\documentclass[12pt]{article}
\usepackage{epsf}
\title{ {\bf
CP violation in the inclusive $b\rightarrow s g $ decay in the framework of
multi-Higgs doublet models}}
\author{\vspace{1cm}\\
{\bf A. Goksu}
        \thanks{E-mail address:
        agoksu@metu.edu.tr}\,\,\, ,
{\bf E. O. Iltan}
        \thanks{E-mail address:
        eiltan@heraklit.physics.metu.edu.tr}\,\,\,  and 
{\bf L. Solmaz}
        \thanks{E-mail address:
        lsolmaz@photon.physics.metu.edu.tr}
 \\
        Physics Department, Middle East Technical University \\
        Ankara, Turkey\\}

\date{}

\begin{document}
\setlength{\baselineskip}{24pt}
\maketitle
\setlength{\baselineskip}{7mm}
\begin{abstract}
We study the decay width and CP asymmetry of the inclusive process 
$b\rightarrow s g$ (g denotes gluon) in the multi Higgs doublet  
models with complex Yukawa couplings, including next to leading QCD
corrections. We analyse the dependencies of the decay width and 
CP asymmetry on the scale $\mu$ and CP violating parameter $\theta$. 
We observe that there exist an enhancement in the decay width and CP 
asymmetry is at the order of $10^{-2}$. 
\end{abstract} 
\thispagestyle{empty}
\newpage
\setcounter{page}{1}
\section{Introduction}
Rare B decays are induced by flavor changing neutral currents (FCNC) at
loop level. Therefore they are phenomenologically rich and provide a 
comprehensive information about the theoretical models and the existing 
free parameters. The forthcoming experiments at SLAC, KEK B-factories, 
HERA-B and possible future accelerators stimulate the study of such decays 
since the large number of events can take place and various branching ratios, 
CP-violating asymmetries, polarization effects, etc., can be measured 
\cite{Bar,Ellis1}. 

Among B decay modes, inclusive $b\rightarrow s g$  is interesting since it 
is theoretically clean and sensitive to new physics beyond the SM, like two 
Higgs doublet model (2HDM) \cite{Glashow}, minimal supersymmetric 
Standard model (MSSM) \cite{Misiak1,Abel}, etc.  

There are various studies on this process in the literature. The Branching 
ratio ($Br$) of $b\rightarrow s g$ decay in the SM is $Br(b\rightarrow s g)
\sim 0.2 \%$ for on-shell gluon \cite{Gao}. This ratio can be enhanced with
the addition of QCD corrections or by taking into account the extensions of 
the SM. The enhanced  $Br (b\rightarrow s g)$ is among the possible 
explanations for the semileptonic branching ratio $B_{SL}$ and the average 
charm multiplicity . The theoretical predictions of $B_{SL}$ 
\cite{Bagan} are slightly different than the experimental measurements 
obtained at the $\Upsilon(4S)$ and $Z^0$ resonance \cite{Yamamoto}. Further 
the measured charm multiplicity $\eta_c$ is smaller than the theoretical 
result. The enhancement of $Br(B\rightarrow X_{no\,charm})$ and therefore 
$Br (b\rightarrow s g)$ rate would explain the missing charm and 
$B_{SL}$ problem \cite{Kagan}. Further, $Br(B\rightarrow \eta^{\prime} X_s)$ 
reported by CLEO \cite{Browder} stimulates to study on the enhancement of 
$Br (b\rightarrow s g)$.

In \cite{Chao1,Kagan2}, the enhancement of $Br\,(b\rightarrow s g)$ was
obtained less than one order compared to the SM case in the framework of 
the 2HDM  (Model I and II)  for $m_{H^{\pm}}\sim 200\, GeV$ and 
$tan\,\beta\sim 5$. The possibility of large $Br$ in the supersymmetric
models was studied in \cite{Berto}. In \cite{Zhen} $Br$ was calculated 
in the model III and the prediction of the enhancement, at least one 
order larger compared to the SM one, makes it possible to describe 
the results coming from experiments \cite{Kagan}. In the case of time-like 
gluon, namely $b\rightarrow s g^*$ decay, $Br$ should be consistent with 
the CLEO data \cite{CLEO1}
\begin{eqnarray} 
Br\,(b\rightarrow s g^*) < 6.8\,\% 
\label{Br1}
\end{eqnarray}   
and in \cite{Zhen}, it was showed that the model III enhancement was not 
contradict with this data for light-like gluon case. 
The calculation of $Br\,(b\rightarrow s g)$ with the addition of next 
to leading logarithmic (NLL) QCD corrections was done in \cite{Greub} and 
it was observed that this ratio enhanced by more than a factor of 2.   

CP violating asymmetry ($A_{CP}$) is another physical parameter which can
give strong clues for the physics beyond the SM. The source of CP violating 
effects in the SM are complex Cabbibo-Cobayashi-Maskawa (CKM) matrix 
elements. $A_{CP}$ for the inclusive $b\rightarrow s g$ decay vanishes  in 
the SM and this forces one to go beyond the SM to check if a measurable 
$A_{CP}$ is obtained.     

In this work, we study the decay width $\Gamma$ and $A_{CP}$ of 
$b\rightarrow s g$ decay in the 3HDM and model III version of 2HDM. In 
these models, it is possible to enhance $\Gamma$ and to get a measurable 
$A_{CP}$. Since the Yukawa couplings for new physics can be chosen complex 
and the addition of NLL corrections \cite{Greub} brings additional complex 
quantities into the amplitude, theoretically, it is possible to get a 
considerable $A_{CP}$, at the order of the magnitude $2\,\%$. This effect 
is due to new physics beyond the SM, 3HDM and model III in our case. 

The paper is organized as follows:
In Section 2, we give a brief summary of the model III and $3HDM(O_2)$ 
and present the expressions appearing in the calculation  of the decay width 
of the inclusive  $b\rightarrow s g$ decay. Further we calculate the 
CP asymmetry $A_{CP}$ of the process. Section 3 is devoted to discussion 
and our conclusions. 
\section{The inclusive process $b\rightarrow s g$ in the framework of the 
multi Higgs doublet models} 

In this section, we study NLL corrected $b\rightarrow s g$ decay width and 
the CP violating effects in the framework of the multi Higgs doublet models 
(model III version of 2HDM and 3HDM)  

In the SM and model I and II 2HDM, the flavour changing neutral current at 
tree level is forbidden. However, they are permitted in the general 
2HDM, so called model III with new parameters, i.e. Yukawa 
couplings. The Yukawa interaction in this general case reads as 
\begin{eqnarray}
{\cal{L}}_{Y}&=&\eta^{U}_{ij} \bar{Q}_{i L} \tilde{\phi_{1}} U_{j R}+
\eta^{D}_{ij} \bar{Q}_{i L} \phi_{1} D_{j R}+
\xi^{U\,\dagger}_{ij} \bar{Q}_{i L} \tilde{\phi_{2}} U_{j R}+
\xi^{D}_{ij} \bar{Q}_{i L} \phi_{2} D_{j R} + h.c. \,\,\, ,
\label{lagrangian}
\end{eqnarray}
where $L$ and $R$ denote chiral projections $L(R)=1/2 (1\mp \gamma_5)$,
$\phi_{k}$, for $k=1,2$, are the two scalar doublets, $Q_{i L}$ are quark 
doublets, $U_{j R}$ and $D_{j R}$ are quark singlets, $\eta^{U,D}_{ij}$ and 
$\xi^{U,D}_{ij}$ are the matrices of the Yukawa couplings. The Flavor 
changing (FC) part of the interaction is given by
\begin{eqnarray}
{\cal{L}}_{Y,FC}=
\xi^{U\,\dagger}_{ij} \bar{Q}_{i L} \tilde{\phi_{2}} U_{j R}+
\xi^{D}_{ij} \bar{Q}_{i L} \phi_{2} D_{j R} + h.c. \,\, .
\label{lagrangianFC}
\end{eqnarray}
The choice of $\phi_1$ and $\phi_2$
\begin{eqnarray}
\phi_{1}=\frac{1}{\sqrt{2}}\left[\left(\begin{array}{c c} 
0\\v+H_0\end{array}\right)\; + \left(\begin{array}{c c} 
\sqrt{2} \chi^{+}\\ i \chi^{0}\end{array}\right) \right]\, ; 
\phi_{2}=\frac{1}{\sqrt{2}}\left(\begin{array}{c c} 
\sqrt{2} H^{+}\\ H_1+i H_2 \end{array}\right) \,\, ,
\label{choice}
\end{eqnarray}
and the vacuum expectation values,  
\begin{eqnarray}
<\phi_{1}>=\frac{1}{\sqrt{2}}\left(\begin{array}{c c} 
0\\v\end{array}\right) \,  \, ; 
<\phi_{2}>=0 \,\, ,
\label{choice2}
\end{eqnarray}
allows us to carry the information about new physics in the doublet $\phi_2$. 
Further, we take $H_{1}$, $H_{2}$ as the mass eigenstates $h_{0}$, $A_{0}$ 
respectively. Note that, at tree level, there is no mixing among CP even 
neutral Higgs particles, namely  the SM one, $H_0$, and beyond, $h_{0}$. 

In eq.(\ref{lagrangianFC}) the couplings  $\xi^{U,D}$ for the FC charged 
interactions are 
\begin{eqnarray}
\xi^{U}_{ch}&=& \xi_{neutral} \,\, V_{CKM} \nonumber \,\, ,\\
\xi^{D}_{ch}&=& V_{CKM} \,\, \xi_{neutral} \,\, ,
\label{ksi1} 
\end{eqnarray}
where  $\xi^{U,D}_{neutral}$ 
is defined by the expression
\begin{eqnarray}
\xi^{U (D)}_{N}=(V_{R(L)}^{U (D)})^{-1} \xi^{U,(D)} V_{L(R)}^{U (D)}
\,\, .
\label{ksineut}
\end{eqnarray}
where $\xi^{U,D}_{neutral}$ is denoted as $\xi^{U,D}_{N}$. 
Here the charged couplings are  the linear combinations of neutral 
couplings multiplied by $V_{CKM}$ matrix elements (see \cite{alil1} for 
details). 
In the case of the general 3HDM, there is an additional Higgs doublet,
$\phi_3$, and the Yukawa interaction can be written as 
\begin{eqnarray}
{\cal{L}}_{Y}&=&\eta^{U}_{ij} \bar{Q}_{i L} \tilde{\phi_{1}} U_{j R}+
\eta^{D}_{ij} \bar{Q}_{i L} \phi_{1} D_{j R}+
\xi^{U\,\dagger}_{ij} \bar{Q}_{i L} \tilde{\phi_{2}} U_{j R}+
\xi^{D}_{ij} \bar{Q}_{i L} \phi_{2} D_{j R} \nonumber \\
&+&
\rho^{U\,\dagger}_{ij} \bar{Q}_{i L} \tilde{\phi_{3}} U_{j R}+
\rho^{D}_{ij} \bar{Q}_{i L} \phi_{3} D_{j R}
 + h.c. \,\,\, ,
\label{lagrangian3H}
\end{eqnarray}
where $\rho^{U,D}_{ij}$ is the new Yukawa matrix having complex entries, in 
general. The similar choice of Higgs doublets   
\begin{eqnarray}
\phi_{1}=\frac{1}{\sqrt{2}}\left[\left(\begin{array}{c c} 
0\\v+H^{0}\end{array}\right)\; + \left(\begin{array}{c c} 
\sqrt{2} \chi^{+}\\ i \chi^{0}\end{array}\right) \right]\, , 
\nonumber \\ \\
\phi_{2}=\frac{1}{\sqrt{2}}\left(\begin{array}{c c} 
\sqrt{2} H^{+}\\ H^1+i H^2 \end{array}\right) \,\, ,\,\, 
\phi_{3}=\frac{1}{\sqrt{2}}\left(\begin{array}{c c} 
\sqrt{2} F^{+}\\ H^3+i H^4 \end{array}\right) \,\, ,\nonumber
\label{choice3H}
\end{eqnarray}
with the vacuum expectation values,  
\begin{eqnarray}
<\phi_{1}>=\frac{1}{\sqrt{2}}\left(\begin{array}{c c} 
0\\v\end{array}\right) \,  \, ; 
<\phi_{2}>=0 \,\, ; <\phi_{3}>=0\,\,  
\label{choice23H}
\end{eqnarray}
can be done and the information about new physics is carried beyond the SM in 
the last two doublets, $\phi_2$ and $\phi_3$. Further, we take $H_{1}$, $H_{2}$, 
$H_{3}$ and $H_{4}$ as the mass eigenstates $h_{0}$, $A_{0}$, $h'_{0}$, 
$A'_{0}$ where $h'_{0}$, $A'_{0}$ are new neutral Higgs bosons due to the
additional Higgs doublet in the 3HDM (see \cite{eril3}).

The Yukawa interaction for the Flavor 
Changing (FC) part is
\begin{eqnarray}
{\cal{L}}_{Y,FC}=
\xi^{U\,\dagger}_{ij} \bar{Q}_{i L} \tilde{\phi_{2}} U_{j R}+
\xi^{D}_{ij} \bar{Q}_{i L} \phi_{2} D_{j R}
+\rho^{U\,\dagger}_{ij} \bar{Q}_{i L} \tilde{\phi_{3}} U_{j R}+
\rho^{D}_{ij} \bar{Q}_{i L} \phi_{3} D_{j R} + h.c. \,\, ,
\label{lagrangianFC3H}
\end{eqnarray}
where the charged couplings  $\xi_{ch}^{U,D}$ and $\rho_{ch}^{U,D}$ are 
\begin{eqnarray}
\xi^{U}_{ch}&=& \xi_{N} \,\, V_{CKM} \nonumber \,\, ,\\
\xi^{D}_{ch}&=& V_{CKM} \,\, \xi_{N}  \nonumber \,\, , \\
\rho^{U}_{ch}&=& \rho_{N} \,\, V_{CKM} \nonumber \,\, ,\\
\rho^{D}_{ch}&=& V_{CKM} \,\, \rho_{N} \,\, ,
\label{ksi13H} 
\end{eqnarray}
and
\begin{eqnarray}
\xi^{U (D)}_{N}=(V_{R(L)}^{U (D)})^{-1} \xi^{U,(D)} V_{L(R)}^{U (D)}\,\, , 
\nonumber \\
\rho^{U (D)}_{N}=(V_{R(L)}^{U (D)})^{-1} \rho^{U,(D)} V_{L(R)}^{U (D)}
\,\, .
\label{ksineut3H}
\end{eqnarray}
Since there exist additional charged Higgs particles, $F^{\pm}$, and 
neutral Higgs bosons $h^{\prime\, 0}$, $A^{\prime\, 0}$ in the 3HDM, 
we introduce a new global $O(2)$ symmetry in the Higgs sector, considering 
three Higgs scalars as orthogonal vectors in a new space, which we call Higgs 
flavor space and we denote the Higgs flavor index by  "$m$", where $m=1,2,3$.
The transformation reads  
\begin{eqnarray}
\phi'_{1}&=&\phi_{1}\nonumber \,\,,\\
\phi'_{2}&=&cos\,\alpha\,\, \phi_2+sin\,\alpha\,\, \phi_3 \,\, , \nonumber \\
\phi'_{3}&=&-sin\,\alpha\,\, \phi_2 + cos\,\alpha\,\, \phi_3\,\,,
\label{trans}
\end{eqnarray}
where $\alpha$ is the global parameter, which represents a rotation of 
the vectors $\phi_2$ and $\phi_3$ along the axis that $\phi_1$ lies, 
in the Higgs flavor space. This symmetry ensures that the new particles are 
mass degenerate with their counterparts existing in model III 
(see \cite {eril3} for details). Further the Yukawa Lagrangian 
(eq.(\ref{lagrangian3H})) is invariant under this transformation if the Yukawa 
matrices satisfy the expressions
\begin{eqnarray}
\bar{\xi}^{\prime U(D)}_{ij}&=& \bar{\xi}^{U (D)}_{ij} cos\, \alpha+
\bar{\rho}^{U(D)}_{ij} sin\, \alpha\,\, \nonumber ,\\
\bar{\rho}^{\prime U (D)}_{ij}&=&-\bar{\xi}^{U (D)}_{ij} sin\, \alpha+
\bar{\rho}^{U (D)}_{ij} cos\, \alpha \,\, .
\label{yuktr}
\end{eqnarray}
and we get 
\begin{eqnarray}
(\bar{\xi}^{\prime U(D)})^+ \bar{\xi}^{\prime U (D) } +
(\bar{\rho}^{\prime U (D)})^+\bar{\rho}^{\prime U (D) }=
(\bar{\xi}^{U (D)})^+\bar{\xi}^{U (D)} +
(\bar{\rho}^{U (D)})^+ \bar{\rho}^{U (D) }\,\, . 
\label{yukinv}
\end{eqnarray}
Therefore, it possible to parametrize the Yukawa matrices $\bar{\xi}^{U(D)}$ 
and $\bar{\rho}^{U(D)}$ as : 
\begin{eqnarray}
\bar{\xi}^{U (D)}=\bar{\epsilon}^{U(D)} cos\,\theta \nonumber \,\, ,\\
\bar{\rho}^{U}=\bar{\epsilon}^{U} sin\,\theta \nonumber \,\, ,\\ 
\bar{\rho}^{D}=i \bar{\epsilon}^{D} sin\,\theta \,\, ,
\label{yukpar}
\end{eqnarray}
where $\bar{\epsilon}^{U(D)}$ are real matrices satisfy the equation 
\begin{eqnarray}
(\bar{\xi}^{\prime U(D)})^+ \bar{\xi}^{\prime U (D) } +
(\bar{\rho}^{\prime U (D)})^+\bar{\rho}^{\prime U (D) }=
(\bar{\epsilon}^{U(D)})^T \bar{\epsilon}^{U(D)} 
\label{yukpareq}
\end{eqnarray}
and the angle $\theta$ is the source of CP violation. Here $X^{U(D)} =
\sqrt{\frac{4 G_F}{\sqrt{2}}}\, \bar{X}^{U(D)}$ with $X=\xi,\,\rho,\,\epsilon$ and  $T$ denotes transpose 
operation. In eq. (\ref{yukpar}),  we take $\bar{\rho}^{D}$ complex to carry 
all CP violating effects in the third Higgs scalar. 

Now, we would like to continue the study of the inclusive process 
$b\rightarrow s g$. Our starting point is the recent calculation of NLL 
corrected decay width \cite{Greub}
\begin{eqnarray}
\Gamma (b\rightarrow s g)= \Gamma^{D} + \Gamma^{brems}\, ,
\label{DWidth}
\end{eqnarray}
where 
\begin{eqnarray}
\Gamma^{D}= c_1\,|D|^2\, ,
\label{DWidthD}
\end{eqnarray}
with 
\begin{eqnarray}
D&=&C_8^{0,eff}+\frac{\alpha_s}{4\,\pi} \{ C_8^{1,eff}-\frac{16}{3} 
C_8^{0,eff} + C_1^{0}\, (l_1 \, ln \frac{m_b}{\mu} + r_1) 
\nonumber \\ 
&+& C_2^{0}\, (l_2\, ln \frac{m_b}{\mu} + r_2)+ C_8^{0,eff} ((l_8+8+\beta_0) 
\, ln\frac{m_b}{\mu}+r_8) \}\, ,
\label{DWidthD2}
\end{eqnarray}
and $\Gamma^{brems}$ is the result for the finite part of bremsstrahlung
corrections 
\begin{eqnarray}
\Gamma^{brems}=c_{2} \int dE_{q} dE_{r} \, (\tau^+_{11} + \tau^+_{22}+
\tau^-_{22}+\tau^+_{12}+\tau^+_{18}+\tau^+_{28}+\tau^-_{28})\, ,
\label{DWidthbr}
\end{eqnarray}
where 
\begin{eqnarray}
\tau^+_{11}&=& 48\, \hat{C}^2_1 \, |\bar\Delta i_{23}|^2\, m_b^2\,(m_b^2-2\, 
E_q\, E_r)\nonumber \,\, ,\\
\tau^+_{22}&=& \frac{56}{3}\, \hat{C}^2_2 \, |\bar\Delta i_{23}|^2\, m_b^2\,
(m_b^2-2\, E_q\, E_r)\nonumber \,\, ,\\
\tau^-_{22}&=& 24\, \hat{C}^2_2 \, |\bar\Delta i_{17}|^2\, 
m_b \,(16\, m_b \,E_q^2-16 \,E_q^2\,E_r- 8 m_b^2\, E_q + 6\,m_b\, E_q\,E_r+
m_b^3)
\nonumber \,\, ,\\
\tau^+_{12}&=& 32\, \hat{C}_1\,\hat{C}_2 \, 
|\bar\Delta i_{23}|^2 \,m_b^2\,(m_b^2-2\, E_q\, E_r) \nonumber \,\, ,\\
\tau^+_{18}&=& 256\, \hat{C}_1 \, Re[C_8^{0\,eff\,*}\, 
\bar\Delta i_{23}]\,m_b^2\,E_q\, E_r \nonumber \,\, ,\\
\tau^+_{28}&=& \frac{16\,\,56}{3}\, \hat{C}_2 \, Re[C_8^{0\,eff\,*}\, 
\bar\Delta i_{23}]\,m_b^2\,E_q\, E_r \nonumber \,\, ,\\
\tau^-_{28}&=& -96\,\hat{C}_2 \, 
Re[ C^{0,eff\,*}_8 \,\bar\Delta i_{17}] \,m_b^4\, (m_b\, (E_q + E_r) -
2\,( E^2_q + E_r^2 + E_q\, E_r)\nonumber \\ 
&+& 4\,\frac{E_q\, E_r\,(E_q+E_r)}{m_b}]/
(E_q\, E_r)
\label{taudefn}
\end{eqnarray}
Here $\hat{C}_1=\frac{1}{2} C_1^0$ and $\hat{C}_2=C_2^0-\frac{1}{6} C_1^0$, 
and $c_1=\frac{\alpha_s \, m_b^5}{24\, \pi^4} 
|G_F\,V_{tb} V^*_{ts}|^2$ and $c_2=\frac{|G_F\,V_{tb} V^*_{ts}|^2\, 
\alpha^2_s }{96\,\, 64 \, \pi^2}$. (see \cite{Greub} for details). 
In eqs. (\ref{DWidthD2}) and (\ref{taudefn}) the Wilson coefficients 
$C_8^{0,eff}$ and $C_{1(2)}^{0}$ (eq. (\ref{C12mu})) includes LL corrections 
and new physics effects enter into the expressions through the coefficients 
$C_8^{0,eff}$ and $C_8^{1,eff}$ (see eq. (\ref{parametr})). The symbol 
$\eta$ is defined as $\eta=\frac{\alpha_s (m_W)}{\alpha_s (\mu)}$ and
$\beta_0=23/3$. The vectors $a_i,\, h_i',\,e'_i,\, f'_i, \, k'_i, \, l'_i,
\, a_i'$, appearing during QCD corrections, and the Wilson coefficients 
$C^{1,\,\rm eff}_4 (m_W)$, $C^{1,\,\rm eff}_1 (m_W)$ and 
$C^{1,eff}_{8} (m_{W})$, the functions $\bar\Delta i_{17}$ and 
$\bar\Delta i_{23}$ in eqs. (\ref{taudefn}), $r_1, \, r_2,\, r_8$  and 
the numbers $l_1, \, l_2,\, l_8$ in eq. (\ref{DWidthD2}) are given in 
\cite{Greub}.   

Now, we would like to start with the calculation of CP asymmetry for the
inclusive decay underconsideration. The possible sources of CP violation 
in the model III (3HDM) are the complex Yukawa couplings. Our procedure is 
to neglect all Yukawa couplings except $\bar{\xi}^U_{N,tt}$ and 
$\bar{\xi}^D_{N,bb}$ ($\bar{\epsilon}^U_{N,tt}$ and 
$\bar{\epsilon}^D_{N,bb}$) (see eqs. (\ref{yukpar}, \ref{yukpareq}) and 
Discussion section) in the model III ($3HDM(O_2)$). Therefore, in the model 
III ($3HDM(O_2)$), only the combination 
$\bar{\xi}^U_{N,tt} \bar{\xi}^D_{N,bb}$ 
($\bar{\epsilon}^U_{N,tt} \bar{\epsilon}^D_{N,bb}$) is responsible for 
$A_{CP}$.  Using the definition of $A_{CP}$ 
\begin{eqnarray}
A_{CP}=\frac{\Gamma(b\rightarrow s g)-
\Gamma(\bar{b}\rightarrow \bar{s} g )}
{\Gamma(b\rightarrow s g )+
\Gamma(\bar{b}\rightarrow \bar{s} g )},\,\,
\label{ACP1}
\end{eqnarray}
we get 
\begin{equation}
A_{CP}=Im[\bar\xi^{D}_{N,bb}] \,\frac{\Omega^{D}+\Omega^{br}}
{\Lambda^{D}+\Lambda^{br}}\, \, ,
\label{ACP2}
\end{equation}
in the model III where 
$\Omega^{D (br)}$ and $\Lambda^{D (br)}$ are the contributions coming from 
D-part (bremsstrahlung-part) and they read as 
\begin{eqnarray}
\Omega^{D}&=& \frac{\alpha_{s}}{\pi}\, c_{1}\, A_{7}\, Im [A_{5}] 
\nonumber \,\, , \\
\Omega^{br}&=& 2\, c_{2} \int dE_{q} dE_{r} (B_{5}\,Im [\bar\Delta_{23}]+
B_{6}\,Im [\bar\Delta i_{17}]) \nonumber \,\, , \\
\Lambda^{D}&=& 2\, c_{1} \{  |A_{6}|^2 + |\bar\xi^{D}_{N,bb}|^2 \,  
|A_{7}|^2 + 2\,A_7\, Re[\xi^{D}_{N,bb}]\, Re[A_6] \} \nonumber \,\, , \\  
\Lambda^{br}&=& 2\, c_{2} \int dE_{q}\, dE_{r} \{ B_{4} + 
Re[\bar\xi^{D}_{N,bb}]\, (B_{5}\, Re[\bar\Delta i_{23}] + B_{6}\, 
Re[\bar\Delta i_{17}]) \}\, .
\label{OmLam}
\end{eqnarray}
The functions $A_{5,6,7}$ and $B_{4,5,6}$ are defined as
\begin{eqnarray}
A_{5} &=& (C_1^{0} (\mu)\, [l_{1}+ln[\frac{m_{b}}{\mu}] + r_{1}] + 
C_2^{0}(\mu)\, [l_{2} + ln[\frac{m_{b}}{\mu}] + r_{2}]) \nonumber
% \\  &+& A_{3}\, [(l_{8} + 8 +\beta_{0}) \,ln[\frac{m_{b}}{\mu}] + r_{8} -
%\frac{16}{3} ] 
\nonumber \,\, , \\
A_{6} &=& (\eta^{14/23} A_{1}+A_{3}) + \frac{\alpha_{s}(\mu)}{4 \pi}\, 
[A_4 + \chi A_1-\frac{16}{3} \eta^{14/23}\, A_1+A_3\nonumber \\ 
&+& (\eta^{14/23}\, A_1+A_3) \, [(l_{8}+8+\beta_{0})\, ln[\frac{m_{b}}{\mu}] 
+ r_{8}]+A_5] \nonumber \,\, , \\
A_{7} &=& \eta^{14/23}\, A_{2}\, \{ 1 + \frac{\alpha_{s}(\mu)}{4 \pi}\, 
[\eta^{-14/23} \chi -\frac{16}{3} + (l_{8}+8+\beta_{0})\, 
ln[\frac{m_{b}}{\mu}] + r_{8}] \}\, , 
\label{A567}
\end{eqnarray}
and 
\begin{eqnarray}
B_4&=&B_1+B_2 \, (\eta^{14/23}\, A_1 + A_3)\, Re[\bar\Delta \,i_{23}] + 
B_3\, (\eta^{14/23} A_1 + A_3)\, Re[\bar\Delta i_{17}] \nonumber \,\, , \\ 
B_5&=&B_2\, \eta^{14/23} A_2 \nonumber \,\, , \\
B_6&=& B_{3}\, \eta^{14/23} A_2 \,\, ,
\label{B456}
\end{eqnarray}
$B_{1,2,3}$ appearing in eq. (\ref{B456}) read
\begin{eqnarray}
B_{1}&=&[\tau_{11}^{+}+\tau_{22}^{+}+\tau_{22}^{-}+\tau_{12}^{+}] 
\nonumber \,\, , \\
B_{2} &=& 32\, m_b^2\, E_q\, E_r\, [8\, \hat{C}_1 + \frac {28}{3}\, 
\hat{C}_2] \nonumber \,\, , \\
B_{3}&=& \frac{\tau^-_{28}}{Re[C_8^{0, eff\,*}\, \bar{\Delta}_{i17}]} \,\, .
\label{B23}
\end{eqnarray}
Here we use the parametrizations 
\begin{eqnarray}
C^{0,eff}_{8}(m_W) &=& A_1+\bar\xi^{D}_{N,bb} \,A_2 \nonumber \,\, , \\
C^{0,eff}_{8}(\mu) &=& \eta^{14/23} \, C^{0,eff}_{8} (m_W) + A_3
\nonumber \,\, ,\\
C^{1,eff}_{8}(\mu)&=&A_4 + \chi\, (A_1+\bar\xi^{D}_{N,bb} A_2)\,\, ,
\label{parametr}
\end{eqnarray}
with 
\begin{eqnarray}
A_{1} &=& C^{SM}_8 (m_W)+C^{H(1)}_8 (m_W) \nonumber \,\, , \\ 
A_{2} &=& C^{H(2)}_8 (m_W) \nonumber \,\, , \\ 
A_{3} &=& \sum_{i=1}^5 h_i^\prime \,\eta^{a^\prime_i} \,
C^{0}_2 (m_W) \nonumber \,\, , \\
A_{4} &=& \eta^{37/23} C^{1,eff}_{8} (m_{W})+ \sum_{i=1}^8 
( e'_i \,\eta \,C^{1,\,\rm eff}_4 (m_W) + (f'_i + k'_i \eta) \, 
C^{0}_2 (m_W) 
\nonumber \\ &+& 
l'_i \,\eta \, C^{1,\,\rm eff}_1 (m_W) ) \eta^{a_i} \,\, ,
\label{A1234}
\end{eqnarray}
and the Wilson coefficients 
\begin{eqnarray}
C_8^{SM}(m_W)&=&-\frac{3 x^2}{4(x-1)^4} \ln x+
\frac{-x^3+5 x^2+2 x}{8 (x-1)^3}\nonumber \, \, , \\ 
C^{H(1)}_8(m_W)&=&\frac{1}{m_{t}^2} \, |\bar{\xi}^{U}_{N,tt}|^2 \, G_{1}(y_t) 
\nonumber \,\, , \\
C^{H(2)}_8(m_W)&=&\frac{1}{m_t m_b} \, (\bar{\xi}^{*U}_{N,tt}) \, 
G_{2}(y_t)\, ,
\label{CH1H2}
\end{eqnarray}
with 
\begin{eqnarray}
G_{1}(y_t)&=& \frac{y_t(-y_t^2+5y_t+2)}{24 (y_t-1)^3}+\frac{-y_t^2} 
{4(y_t-1)^4} \, \ln y_t
\nonumber  \,\, ,\\ 
G_{2}(y_t)&=& \frac{y_t(y_t-3)}{4 (y_t-1)^2}+\frac{y_t} {2(y_t-1)^3} \, 
\ln y_t 
\label{G1G2}
\end{eqnarray}
The LL corrected Wilson coefficients $C^{0}_{1}$ and $C^{0}_{2}$ are
\begin{eqnarray}
C^{0}_{1}(\mu)&=&(\eta^{6/23}-\eta^{-12/23})\, C^{0}_{2}(m_W) 
\nonumber \,\, , \\ 
C^{0}_{2}(\mu)&=&(\frac{2}{3}\eta^{6/23}+\frac{1}{3}\eta^{-12/23}) 
C^{0}_{2}(m_W) \, ,
\label{C12mu}
\end{eqnarray}
and 
\begin{eqnarray}
C^{0}_{2}(m_W)&=& 1 \nonumber \,\, , \\ 
C^{0}_{1}(m_W)&=& 0 \, .
\label{C12mW}
\end{eqnarray}
In eq. (\ref{A567}) the parameter $\chi$ is given by   
\begin{equation}
\chi=6.7441\, (\eta^{37/23}-\eta^{14/23})\, .
\label{chi}
\end{equation}

In our calculations we take only $\bar{\xi}^{D}_{N, bb}$ complex, 
$\bar{\xi}^{D}_{N, bb}=|\bar{\xi}^{D}_{N, bb}|\, e^{i\,\theta}$, where 
$\theta$ is the CP violating parameter which is restricted by the 
experimental upper limit of the neutron electric dipole moment eq. 
(\ref{neutrdip}). For $3HDM(O_2)$, it is necessary to make the following
replacements:
\begin{eqnarray}
\bar{\xi}^{U}_{N, tt} &\rightarrow& \bar{\epsilon}^{U}_{N, tt}
\nonumber \,\, ,\\
Im[\bar{\xi}^{D}_{N, bb}] &\rightarrow& \bar{\epsilon}^{D}_{N, bb}\, 
sin^2\,\theta \nonumber \,\, ,\\
Re[\bar{\xi}^{D}_{N, bb}] &\rightarrow& \bar{\epsilon}^{D}_{N, bb}\, 
cos^2\,\theta \nonumber \,\, ,\\
|\bar{\xi}^{D}_{N, bb}|^2 &\rightarrow& (\bar{\epsilon}^{D}_{N, bb})^2\, .
\label{ksirhopar}
\end{eqnarray}
\section{Discussion}
The general 3HDM model contains large number of free parameters, such as 
masses of charged and neutral Higgs bosons, complex Yukawa matrices, 
$\xi_{ij}^{U,D}$, $\rho_{ij}^{U,D}$ with quark family indices $i,j$. 
First, a new global $O(2)$ symmetry is introduced in the Higgs flavor space   
to connect the Yukawa matrices in the second and third doublet  
and to keep the masses of new charged (neutral) Higgs particles 
in the third doublet degenerate to the ones in the second doublet
\cite{eril3}. Second, the Yukawa couplings, which are entries of Yukawa 
matrices, is restricted using the experimental measurements, namely, 
$\Delta F=2$ mixing, the $\rho$ parameter \cite{Soni} and the CLEO 
measurement \cite{CLEO2}, 
\begin{eqnarray}
Br (B\rightarrow X_s\gamma)= (3.15\pm 0.35\pm 0.32)\, 10^{-4} \,\, .
\label{br2}
\end{eqnarray}
The constraints for the FC couplings from $\Delta F=2$ processes and $\rho$ 
parameter for the model III were investigated without QCD corrections 
\cite{Soni} and the following predictions are reached:
\begin{eqnarray}
\lambda_{uj}&=&\lambda_{dj} <<  1\,,\,\, i,j=1,2,3 \nonumber\,\, ,
\end{eqnarray}
where $u(d)$ is up (down) quark and  $i,j$ are the generation numbers
and  further 
\begin{eqnarray}
\lambda_{bb}\,\,, \lambda_{sb} >> 1 \,\, \hbox{and} \,\, 
\lambda_{tt}\,\,, \lambda_{ct} << 1\,\, .
\label{lambda}
\end{eqnarray}
In the analysis, the ansatz proposed by Cheng and Sher, 
\begin{eqnarray}
\xi_{ij}^{U D}=\lambda_{ij}\sqrt{\frac{m_i m_j}{v}}\,\, ,
\label{ansatz}
\end{eqnarray}
is used. Respecting these constraints and using the measurement
by the CLEO \cite{CLEO2} Collaboration we neglect all Yukawa couplings except 
$\bar{\xi}^{U}_{N,tt}$, $\bar{\xi}^{D}_{N,bb}$ in the model III. In 
$3HDM(O_2)$, the same restrictions are done by taking into account only the 
couplings $\bar{\epsilon}^{U}_{N,tt}$ and $\bar{\epsilon}^{D}_{N,bb}$.  

This section is devoted to the study of the CP parameter $sin\theta$ and 
the scale $\mu$ dependencies of the decay width $\Gamma$ and CP asymmetry 
of $A_{CP}$ for the inclusive decay $b\rightarrow s g$, in the framework 
of the model III and $3HDM(O_2)$. In our analysis, we restrict the 
parameters  $\theta$, $\bar{\xi}^{U}_{N,tt}$ and $\bar{\xi}^{D}_{N bb}$
($\bar{\epsilon}^{U}_{N,tt}$ and $\bar{\epsilon}^{D}_{N bb}$) 
in the model III ($3HDM(O_2)$), using the constraint for $|C_7^{eff}|$, 
$0.257 \leq |C_7^{eff}| \leq 0.439$, coming from the CLEO
data eq. (\ref{br2}) (see \cite{alil1}). Here $C_7^{eff}$ is the effective 
magnetic dipole type Wilson coefficient for $b\rightarrow s\gamma$ vertex. 
The above restriction allows us to define a constraint region for the 
parameter $\bar{\xi}^{U}_{N,tt}$ ($\bar{\epsilon}^{U}_{N,tt}$) in terms of 
$\bar{\xi}^{D}_{N,bb}$ ($\bar{\epsilon}^{D}_{N,bb}$) and $\theta$ in the 
the model III, ($3HDM(O_2)$). Further, in our numerical calculations we 
respect the constraint for the angle $\theta$, due to the experimental upper 
limit of neutron electric dipole moment, namely  
\begin{eqnarray}
d_n<10^{-25}\hbox{e$\cdot$cm}
\label{neutrdip}
\end{eqnarray}
which places an upper bound on the couplings with the expression  
in the model III ($3HDM(O_2)$): 
$\frac{1}{m_t m_b} (\bar{\xi}^{U}_{N,tt}\,\bar{\xi}^{* D}_{N,bb})\, sin\,
\theta < 1.0$ ($\frac{1}{m_t m_b} (\bar{\epsilon}^{U}_{N,tt}\,
\bar{\epsilon}^{* D}_{N,bb})\, sin^2\,\theta < 1.0$) for 
$m_{H^\pm}\approx 200$ GeV \cite{David}. 

Throughout these calculations, we take the charged Higgs mass
$m_{H^{\pm}}=400\, GeV$, and we use the input values given in Table 
(\ref{input}).  
\begin{table}[h]
        \begin{center}
        \begin{tabular}{|l|l|}
        \hline
        \multicolumn{1}{|c|}{Parameter} & 
                \multicolumn{1}{|c|}{Value}     \\
        \hline \hline
        $m_c$                   & $1.4$ (GeV) \\
        $m_b$                   & $4.8$ (GeV) \\           
        $|V_{tb}\,V^*_{ts}|$    & 0.04 \\
        $m_{t}$             & $175$ (GeV) \\
        $m_{W}$             & $80.26$ (GeV) \\
        $m_{Z}$             & $91.19$ (GeV) \\
        $\Lambda_{QCD}$             & $0.214$ (GeV) \\
        $\alpha_{s}(m_Z)$             & $0.117$  \\
        \hline
        \end{tabular}
        \end{center}
\caption{The values of the input parameters used in the numerical
          calculations.}
\label{input}
\end{table}

\begin{figure}[htb]
\vskip -1.80truein
\centering
\epsfxsize=5in
\leavevmode\epsffile{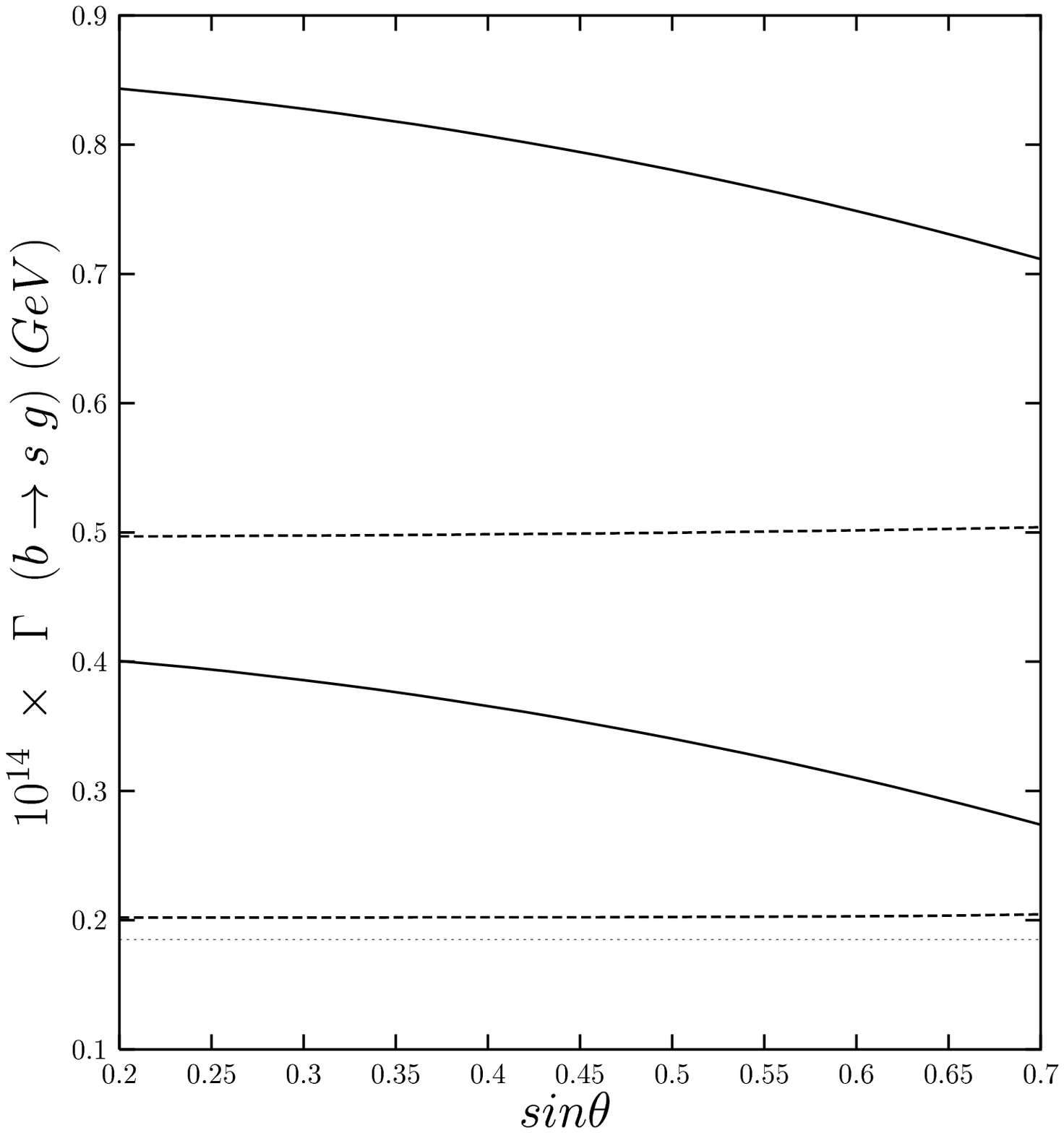}
\vskip -1.8truein
\caption[]{$\Gamma$ as a function of 
$sin\theta$ for $|r_{tb}|=|\frac{\bar{\xi}_{N,tt}^{U}}
{\bar{\xi}_{N,bb}^{D}}| <1$, $\bar{\xi}_{N,bb}^{D}=40\, m_b$ and $\mu=m_b$.  
Here $\Gamma$  is restricted in the region bounded 
by solid (dashed) lines for $C_7^{eff} > 0$ ($C_7^{eff} < 0$), in the model
III. Dotted line represents the SM contribution.}
\label{Gammasin}
\end{figure}

\begin{figure}[htb]
\vskip -1.90truein
\centering
\epsfxsize=5in
\leavevmode\epsffile{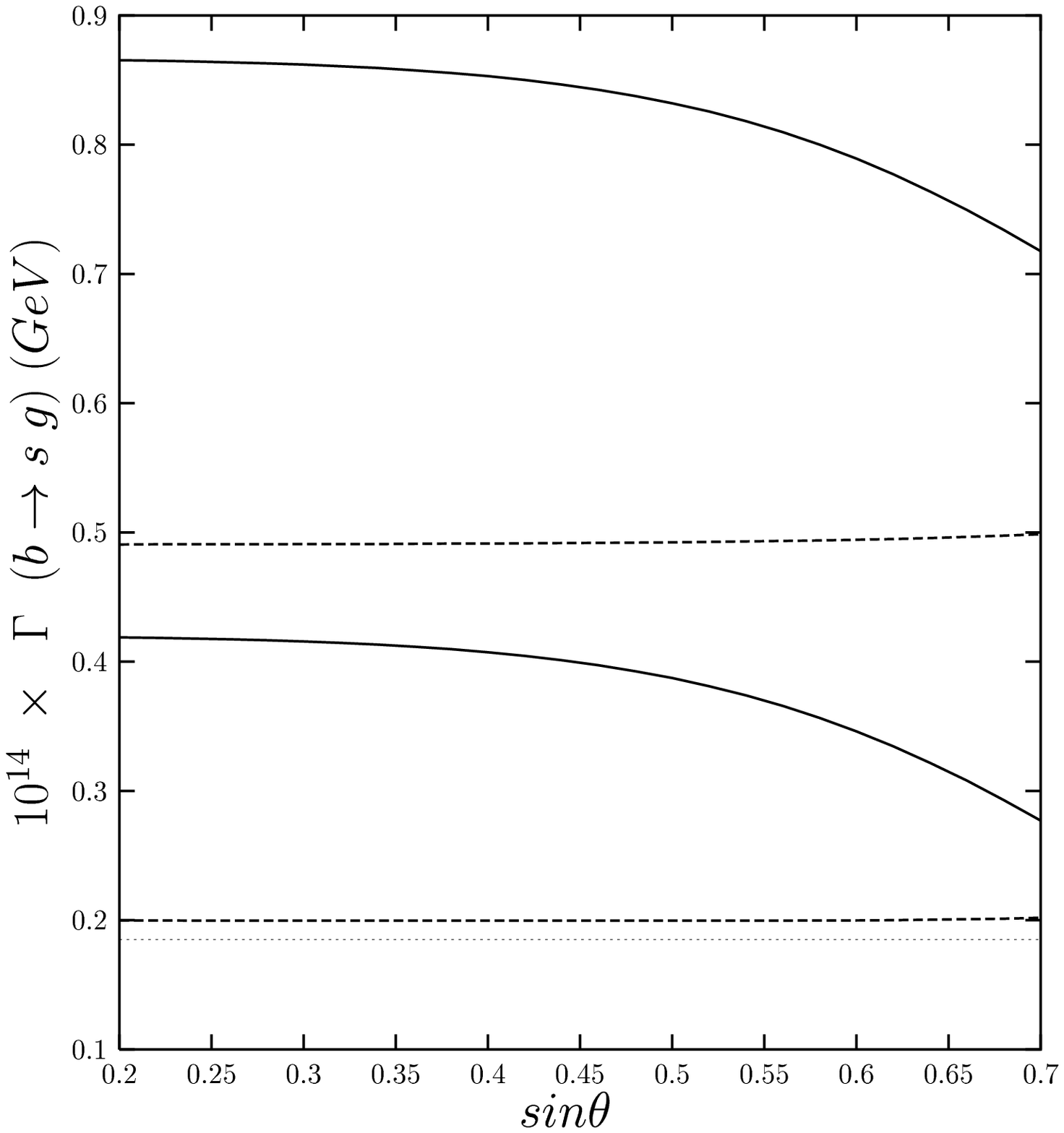}
\vskip -1.9truein
\caption[]{The same as Fig. \ref{Gammasin} but for $3HDM(O_2)$.}
\label{Gammasin3H}
\end{figure}

\begin{figure}[htb]
\vskip -1.9truein
\centering
\epsfxsize=5in
\leavevmode\epsffile{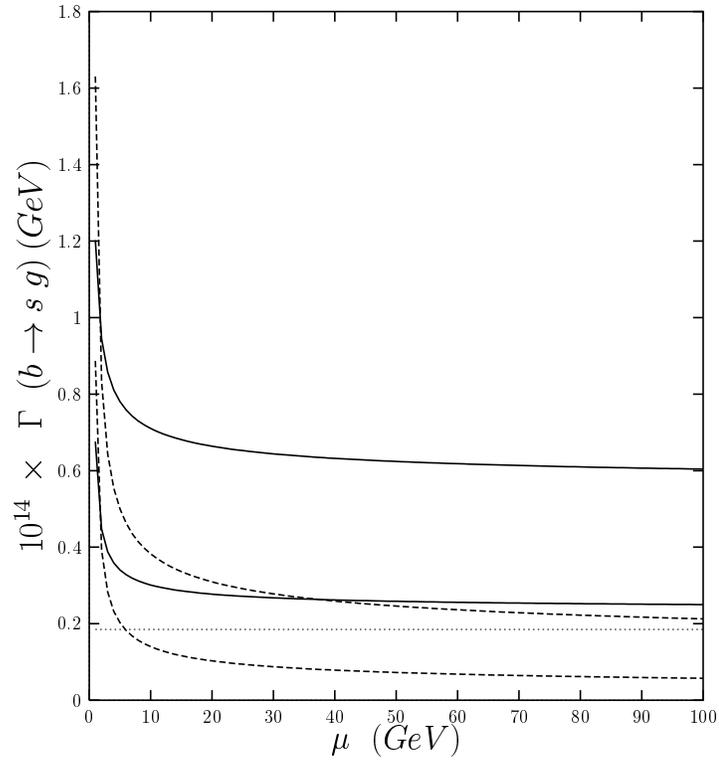}
\vskip -1.9truein
\caption[]{The same as Fig. \ref{Gammasin} but $\Gamma$ as a function of 
$\mu$ for $sin\theta=0.5$.}
\label{Gammamu}
\end{figure}

\begin{figure}[htb]
\vskip -1.9truein
\centering
\epsfxsize=5in
\leavevmode\epsffile{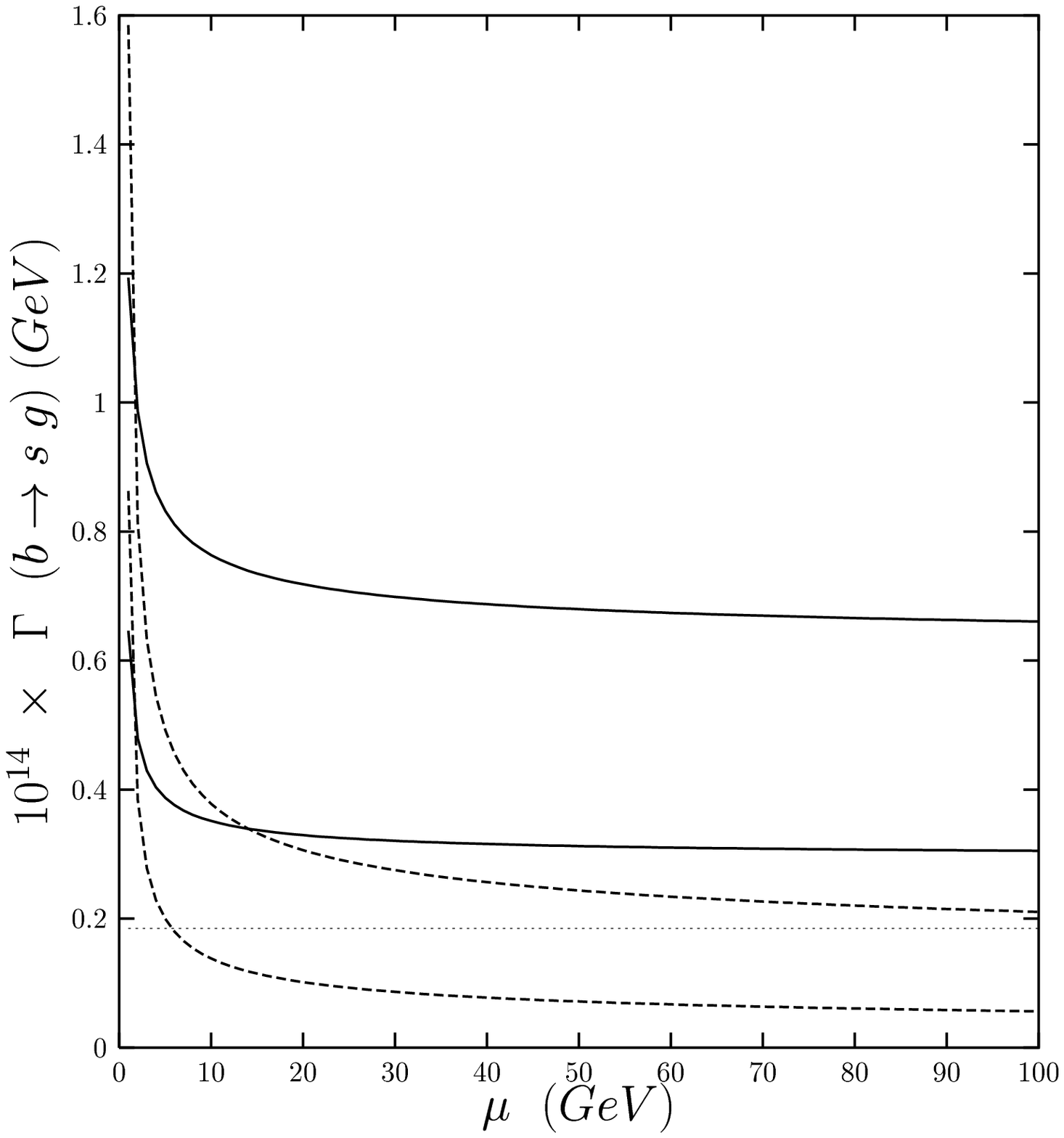}
\vskip -1.9truein
\caption[]{The same as Fig. \ref{Gammamu} but for $3HDM(O_2)$.}
\label{Gammamu3H}
\end{figure}

Fig. \ref{Gammasin} (\ref{Gammasin3H}) is devoted to the $sin\, \theta$ 
dependence of $\Gamma$ for $\mu=m_b$, $\bar{\xi}_{N,bb}^{D}=40\, m_b$ 
($\bar{\epsilon}^{D}_{N,bb}=40\, m_b$) and 
$|r_{tb}|=|\frac{\bar{\xi}_{N,tt}^{U}} {\bar{\xi}_{N,bb}^{D}}| <1$ 
($|\frac{\bar{\epsilon}_{N,tt}^{U}} {\bar{\epsilon}_{N,bb}^{D}}| <1$)
in the model III ($3HDM (O_2)$). Here $\Gamma$ 
is restricted between solid (dashed) lines for $C_7^{eff} > 0$ 
($C_7^{eff} < 0$). As shown in Fig. \ref{Gammasin}, the decay width $\Gamma$ 
can reach $(0.78\pm 0.06)\times 10^{-14}$ in the region 
$0.2\leq sin\,\theta \leq 0.7$ for $C_7^{eff} > 0$ and the possible 
enhancement, a factor of $4.2$ compared to the SM one 
($0.185\pm 0.037)\times 10^{-14}\, GeV$ \cite{Greub} can be reached. 
For $3HDM (O_2)$, the upper range for the decay width $\Gamma$ is 
$(0.79\pm 0.07)\times 10^{-14}$ in the region $0.2\leq sin\,\theta \leq 0.7$ 
for $C_7^{eff} > 0$ and this leads to an enhancement, a factor of $4.3$ 
compared to the SM one.
$\Gamma$ decreases with increasing $sin\theta$ for $C_7^{eff} > 0$ and it 
can get larger values compared to the $C_7^{eff} < 0$ case, in both models. 
The $sin\,\theta$ dependence of $\Gamma$ is weak for $C_7^{eff} < 0$ and for
this case, it takes slightly smaller values in the $3HDM(O_2)$ compared to 
the ones in the model III. In our numerical calculations, we observe that 
the contribution of bremsstrahlung corrections are almost one order smaller 
as a magnitude compared to the rest. Further, the restriction regions for  
$C_7^{eff} > 0$ and $C_7^{eff} < 0$ become more seperated with increasing 
values of the scale $\mu$ and this behaviour is strong in the $3HDM(O_2)$. 
The scale dependence of $\Gamma$ is weak for the values $\mu > 2\, GeV$ 
and almost no dependence is observed for the large values of $\mu$ scale 
for both models. (see Figs. \ref{Gammamu} and \ref{Gammamu3H}). 

\begin{figure}[htb]
\vskip -1.9truein
\centering
\epsfxsize=5in
\leavevmode\epsffile{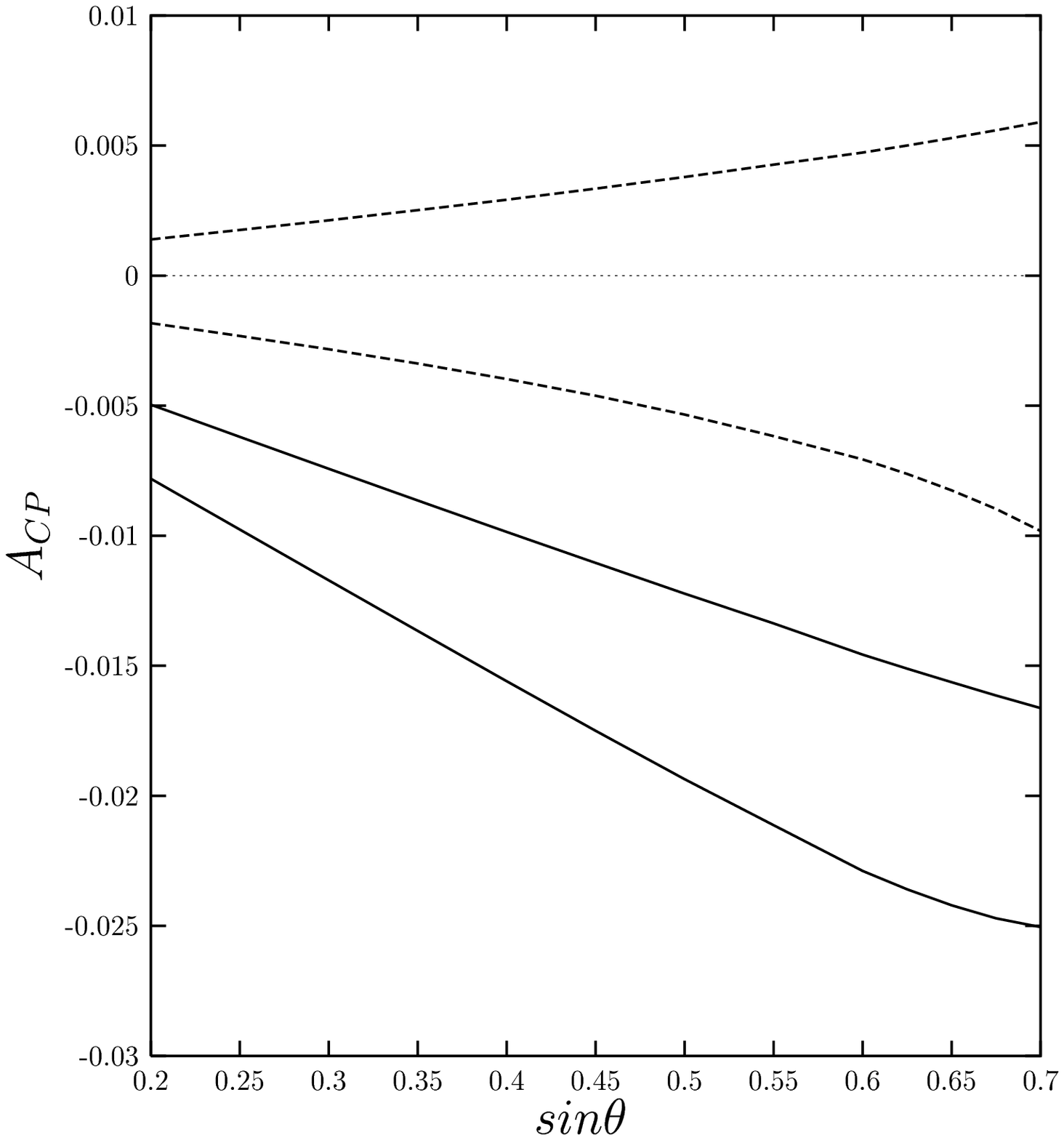}
\vskip -1.9truein
\caption[]{The same as Fig. \ref{Gammasin} but $A_{CP}$ as a function of 
$sin\theta$.}
\label{ACPsin}
\end{figure}

\begin{figure}[htb]
\vskip -1.9truein
\centering
\epsfxsize=5in
\leavevmode\epsffile{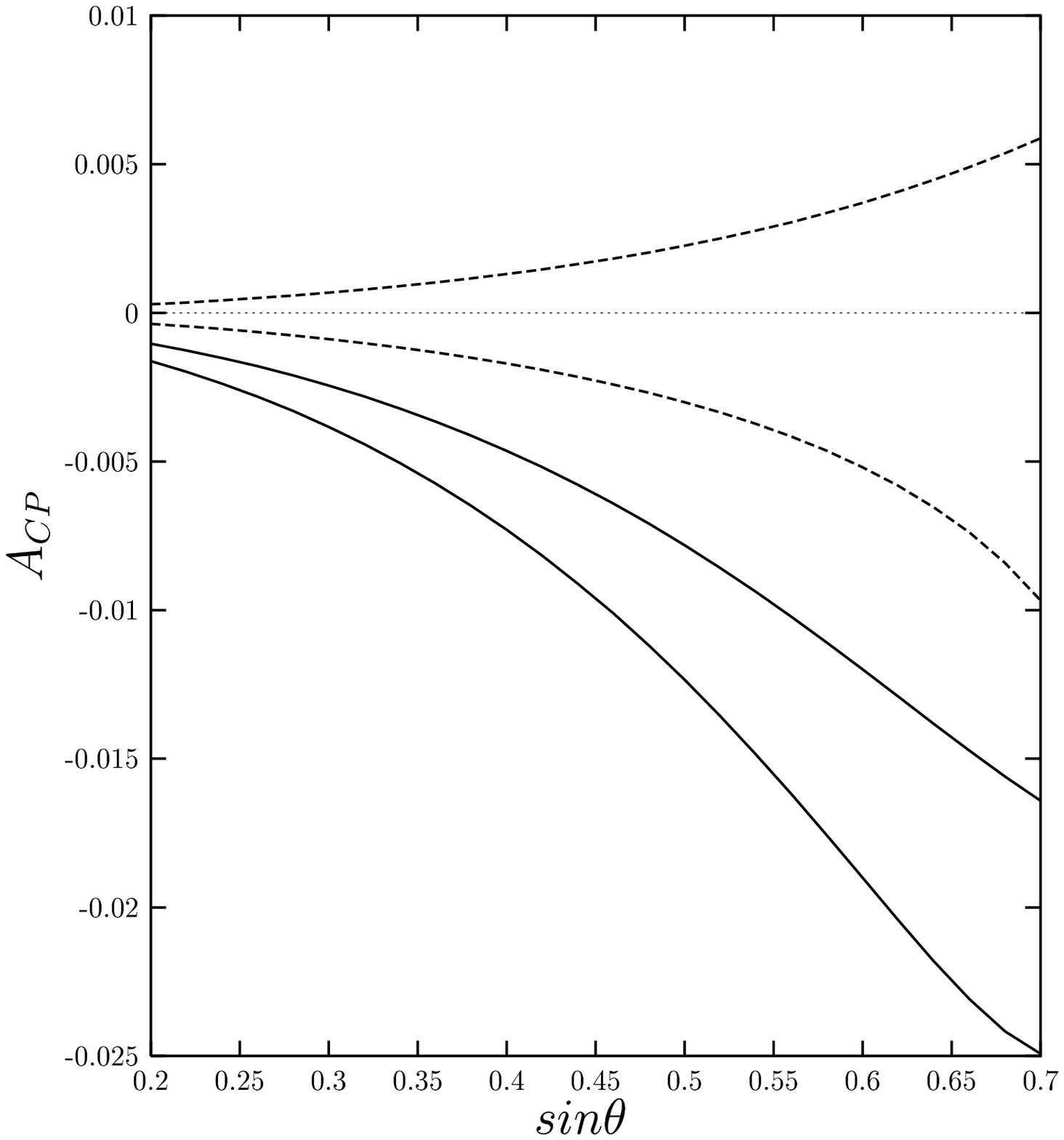}
\vskip -1.9truein
\caption[]{The same as Fig. \ref{Gammasin3H} but $A_{CP}$ as a function of 
$sin\,\theta$.}
\label{ACPsin3H}
\end{figure}

In  Fig. \ref{ACPsin} and \ref{ACPsin3H}, we present the $sin\,\theta$ 
dependence of $A_{CP}$ for $\mu=m_b$, $\bar{\xi}_{N,bb}^{D}=40\, m_b$ 
($\bar{\epsilon}^{D}_{N,bb}=40\, m_b$) 
and $|r_{tb}|<1$ in the model III ($3HDM(O_2)$). Here 
$A_{CP}$ is  restricted in the region bounded by solid (dashed) lines for 
$C_7^{eff} > 0$ ($C_7^{eff} < 0$). As shown in figures, $|A_{CP}|$ reaches 
$2.5\,\%$ for $sin\,\theta=0.7$ and all possible values of $A_{CP}$ are 
negative. However, for $C_7^{eff} < 0$, the allowed region becomes broader 
and $A_{CP}$ can take both signs, even can vanish. For this case, 
$|A_{CP}|$ reaches almost $1\,\%$  as an upper limit in both models. 
Further $A_{CP}$ is more sensitive to $sin\,\theta$ in the $3HDM (O_2)$ 
compared to the model III. 

\begin{figure}[htb]
\vskip -1.9truein
\centering
\epsfxsize=5in
\leavevmode\epsffile{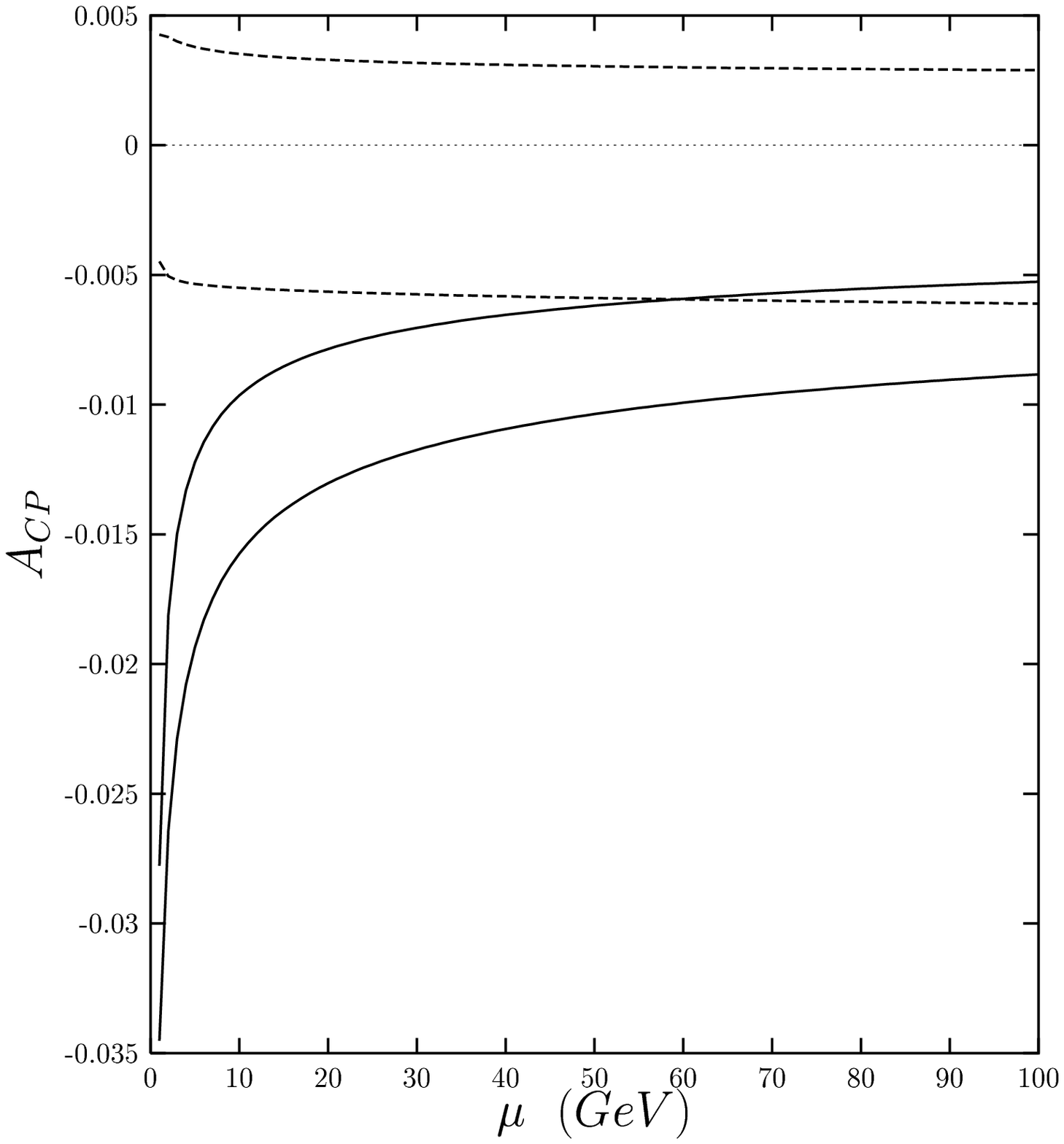}
\vskip -1.9truein
\caption[]{The same as Fig. \ref{Gammamu} but $A_{CP}$ as a function of 
$\mu$.}
\label{ACPmu}
\end{figure}

\begin{figure}[htb]
\vskip -1.9truein
\centering
\epsfxsize=5in
\leavevmode\epsffile{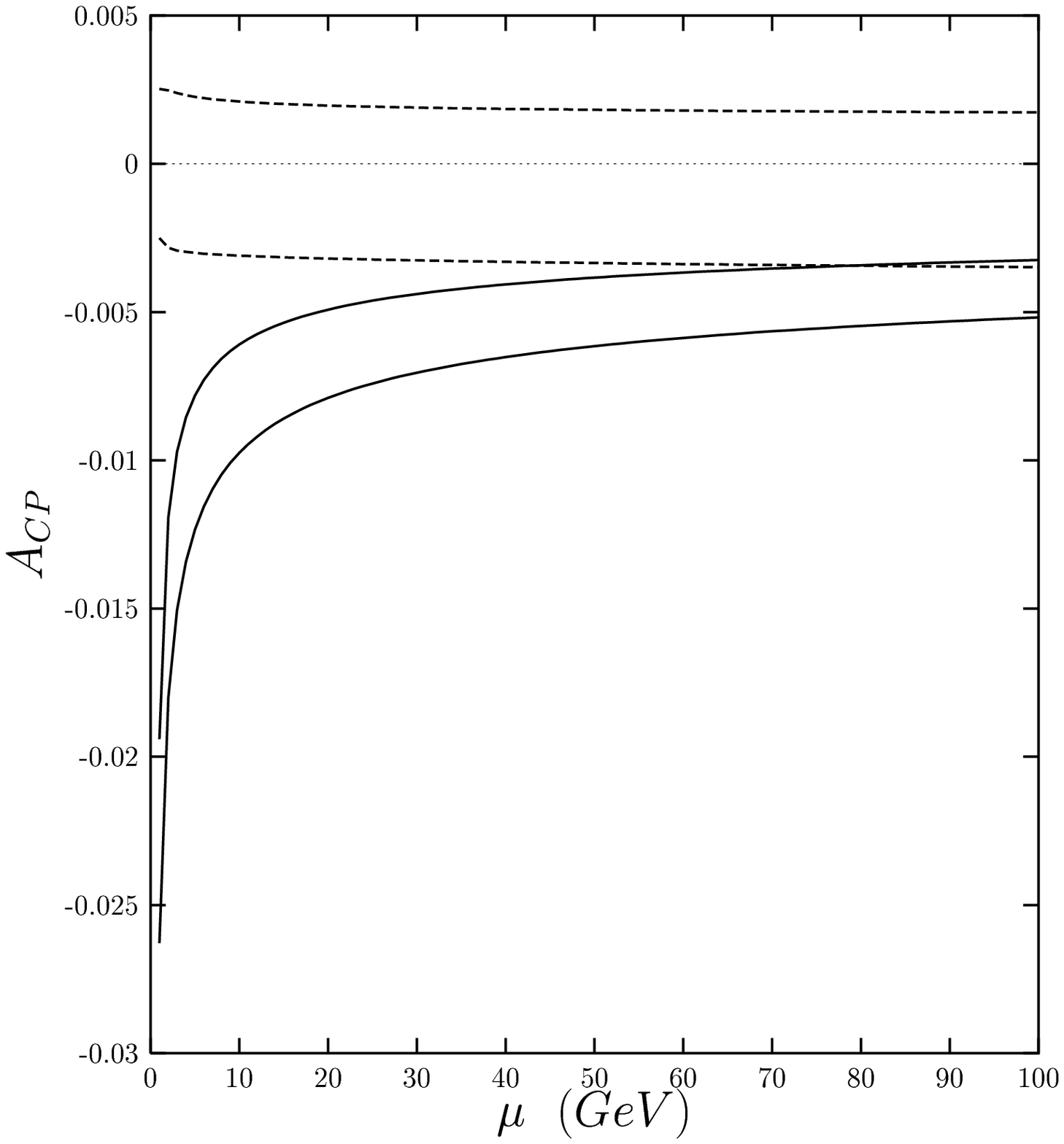}
\vskip -1.9truein
\caption[]{The same as Fig. \ref{Gammamu3H} but $A_{CP}$ as a function of 
$\mu$.}
\label{ACPmu3H}
\end{figure}

Fig. \ref{ACPmu} and \ref{ACPmu3H} represent the scale $\mu$ dependence of 
$A_{CP}$ for $sin\theta=0.5$, $|\bar{\xi}_{N,bb}^{D}|\,\,
(\bar{\epsilon}^{D}_{N,bb})=40\, m_b$ and 
$|r_{tb}|<1$ in both models underconsideration. The scale dependence of 
$A_{CP}$ is also weak for the values $\mu > 2\, GeV$ similar to that of 
$\Gamma$. Here the increasing values of $sin\,\theta$ cause to increase 
the size of restriction region.

At this stage we give the numerical values of $\Gamma$ and $A_{CP}$ for 
$|\bar{\xi}_{N,bb}^D|=40\, m_b$ $(\bar{\epsilon}^{D}_{N,bb}=40\,m_b)$ 
and $\mu=m_b$ in the range $0.2 \leq sin\theta \leq 0.7 $, for model III  
$(3HDM(O_2))$:
\begin{eqnarray}
0.72\, (0.72)\times 10^{-14} \, GeV \leq & \Gamma & \leq 0.84\, (0.86) 
\times 10^{-14}\, GeV\,\, (\hbox{upper boundary}) \,\, for \, C_7^{eff} > 0 
\nonumber \,\, ,\\
0.28 \, (0.28) \times 10^{-14} \, GeV \leq & \Gamma & \leq 0.40\, (0.42) 
\times 10^{-14} \, GeV\,\, (\hbox{lower boundary})\,\, for \, C_7^{eff} > 0 
\nonumber \,\, , \\ 
&\Gamma &=0.50\, (0.48) \times 10^{-14}\, GeV \,\, (\hbox{upper boundary}) 
\,\, for \, C_7^{eff} < 0 
\nonumber \,\, ,\\
&\Gamma &=0.20\, (0.20) \times 10^{-14} \, GeV \,\, (\hbox{lower boundary}) 
\,\, for \, C_7^{eff} < 0
\nonumber \,\, ,\\
\label{Gamnum}
\end{eqnarray}
and
\begin{eqnarray}
0.0080\,(0.0015) \leq &|A_{CP}|& \leq 0.0250\,(0.0250) \,\, 
(\hbox{upper boundary)}\,\, for \, C_7^{eff} > 0 
\nonumber \,\, ,\\
0.0050\,(0.0010) \leq &|A_{CP}|& \leq 0.0170\,(0.0165) \,\, 
(\hbox{lower boundary})\,\, for \, C_7^{eff} > 0 
\nonumber \,\, ,\\
0.0020\,(0.0010) \leq &A_{CP}& \leq 0.0060\,(0.0060) \,\, 
(\hbox{upper boundary})\,\, for \, C_7^{eff} < 0 
\nonumber \,\, ,\\
-0.0100\,(-0.0100) \leq &A_{CP}& \leq -0.0020\,(-0.0010) \,\, 
(\hbox{lower boundary})\,\ for \, C_7^{eff} < 0 
\nonumber \,\, .\\
\label{ACPnum}
\end{eqnarray}
Now we would like to present our conclusions:
\begin{itemize}
\item $\Gamma$ can reach $0.84 \,(0.86)\,\times 10^{-14}$ in the model III
($3HDM(O_2)$) and this is an enhancement a factor of $4$ compared 
to the SM one.
\item A measurable CP asymmetry $A_{CP}$ exists with the addition of NLL
QCD corrections and choice of complex Yukawa coupling 
$\bar{\xi}_{N,bb}^D$ ($\bar{\rho}_{N,bb}^D$ (see section 2)) in the model
III ($3HDM(O_2)$). $|A_{CP}|$ can be obtained at the order of the magnitude 
of $\% \, 2.5$. This physical parameter is coming from the new physics 
effects and it can give strong clues about the physics beyond the SM.   
\end{itemize}
\newpage
\newpage
\end{document}